# Summary and Conclusions of the First DESY Test Beam User Workshop

*J.-H. Arling[1], M.S. Amjad[2], L. Bandiera[3] T. Behnke[1], D. Dannheim[4], R. Diener[1], J. Dreyling-Eschweiler[1], H. Ehrlichmann[1], A. Gerbershagen[4], I.-M. Gregor[1], A. Hayrapetyan[5], H. Jansen[1], J. Kaminski[6], J. Kroll[7], P. Martinengo[4], N. Meyners[1], C. Müntz[8], L. Poley[9], B. Schwenker[10], M. Stanitzki[1]*

## Introduction

On October 5/6, 2017, DESY hosted the first DESY Test Beam User Workshop [1] which took place in Hamburg. Fifty participants from different user communities, ranging from LHC (ALICE, ATLAS, CMS, LHCb) to FAIR (CBM, PANDA), DUNE, Belle-II, future linear colliders (ILC, CLIC) and generic detector R&D presented their experiences with the DESY II Test Beam Facility, their concrete plans for the upcoming years and a first estimate of their needs for beam time in the long-term future beyond 2025. A special focus was also on additional improvements to the facility beyond its current capabilities.

## Current Status

The DESY II Test Beam Facility [2] is operating three independent beam lines providing electrons with a momentum ranging from 1-6 GeV/c. Besides the usual infrastructure like Ethernet, patch panels and gas distribution systems, the facility provides EUDET-type pixel beam telescopes [3] in two of the beam lines as well as the PCMAG 1T solenoid and the BRM 1.4 T Dipole, which are available for the test beam users. The facility is very well used by a world-wide community. The overall assessment is that the facility is running with very high efficiency and meets the needs of the user community.

---

[1] DESY, Notkestr. 85, D-22607 Hamburg, Germany
[2] Department of Physics and Astronomy, UCL Gower Street, London, WC1E 6BT, United Kingdom
[3] INFN Sezione di Ferrara, Via Saragat 1, 44122 Ferrara, Italy
[4] CERN, Route de Meyrin, 1217 Meyrin, Switzerland
[5] II. Phys. Institut, University of Giessen, Heinrich-Buff-Ring 16, D-35392 Giessen, Germany
[6] University of Bonn, Physikalisches Institut, Nußallee 12, D-53115 Bonn, Germany
[7] Institute of Physics of the Czech Academy of Sciences Na Slovance 1999/2, 18221 Prague 8, Czech Republic
[8] Goethe-Universität Frankfurt, Institut für Kernphysik, Max-von-Laue-Straße 1, D-60438 Frankfurt/Main, Germany
[9] DESY, Platanenallee 6, D-15738 Zeuthen, Germany
[10] II. Physikalisches Institut, Universität Göttingen, Friedrich-Hund-Platz 1, D-37077 Göttingen, Germany

A key to the current success is the high beam availability at DESY made possible thanks to the very reliable operation of the DESY II synchrotron and the support available to the users from DESY staff. The running time of the facility, starting typically mid-February and running until the December Christmas break, is considered adequate by most users. Particularly having test beams available in spring and fall of the year is considered vital and complements the test beam availability at the CERN PS and SPS nicely. The DESY II Test Beam Facility has received additional EU funding from the EUDET (FP6), AIDA (FP7) and AIDA2020 (Horizon2020) projects, which also allowed supporting user groups to come to DESY and to perform measurements at the facility.

The EUDET-type pixel beam telescopes are a key instrumentation at DESY and at other test beams worldwide. They are in very high demand. The telescopes are one of the very few examples of beam telescopes that are used by many different users from various experiments. Their success is not only possible thanks to their excellent hardware but also because a central DAQ package (EUDAQ and its successor EUDAQ2) and a reconstruction software package (EUTelescope) are provided, which are maintained by an active developer community currently supported strongly by DESY. The telescopes and other infrastructures like the PCMAG solenoid have been strongly and continuously supported by EU projects starting from EUDET till AIDA2020 [4].

## Additional User Needs

During the workshop, many suggestions to further improve the DESY Test Beam were made, which can be summarized in two main categories: improvements to the provided beams and enhancements to the instrumentation.

The main topic raised by the community was an increase of the particle rate in each beam line, which is currently limited to a few kHz depending on the selected energy. Currently DESY II is running with only one bunch with about $14 \times 10^9$ electrons and a 1 microsecond repetition rate. A way forward would be to increase the rate by injecting more bunches into DESY II, which is currently not possible because of the injection kicker scheme. Such an upgrade would allow to immediately increase the available rate for the users without requiring major changes. Another option requested by the users is to run short bunch trains with a spacing of O(25) ns thereby emulating conditions at colliders like the LHC.

A second issue raised was the intensity per bunch. Increasing this to 20 or 30 tracks per $cm^2$ at the highest possible energy crossing the detector under test would be another priority. This would require using the DESY II primary beam directly which is currently not foreseen.

A third issue was the possibility to have pions and muons of a few GeV for detector tests, which is particularly important for calorimeter tests. This is currently not possibly and would require dumping the DESY II primary beam on a target to produce secondary pions and muons from the pion decay. This could be met with a fourth beam line (see R-Weg Upgrade)

A further point that was raised was to provide a permanently installed tagged photon beam for calorimeter tests. Tagged photon beams have already been produced by individual groups, but there was the desire to have a more permanent setup centrally available. Equally, several groups expressed strong interest to study their detectors with electrons with less than one GeV. Hence a request was made to study the beam line performance between 200 and 1000 MeV and to characterize the available rates. Especially the latter request will be dealt with during the startup of the DESY II synchrotron in 2018.

In terms of instrumentation it is clear that significantly higher rates require new beam telescopes to take advantage of it. Besides a higher rate capability, a precise hit timing for the telescopes was requested by the user community. This can be realized by a dedicated timing layer added to the telescopes or a telescope with new sensors that provide precise time stamping for all the particle hits. Since fast timing is required by several groups also without using a telescope, it was suggested, whether such a system could be provided centrally.  Ideally, such a system would provide a timing resolution at the 20-50 ps level. The idea here is to repeat the success of the EUDET-type telescopes by providing a standardized timing layer instead of many individual solutions. Finally, having a common cold box available at the Test Beam was suggested. This is regularly required, when irradiated sensors are being tested and could reduce the problems individual groups frequently encounter when they are setting up their cooling solutions at DESY.

## Test Beams at DESY 2019-2024

In the next years, especially with the advancement of the high-luminosity upgrades at the LHC and other ongoing projects like e.g. the upcoming detectors at FAIR, the need for test beams will constantly increase. With the Long Shutdown 2 (LS2), there will be no test beam available at CERN for at least two years (2019/2020) and the DESY II Test Beam Facility will be the only place in Europe offering multi-GeV particle beams. From experience during the last shutdown at CERN in 2013, it is already anticipated that the DESY II Test Beam Facility will be very much in demand. Hence for the years 2019 and 2020 there is a strong wish to maximize the running time of DESY II each year, with a start of user operations at the end of January until the Christmas break with a short summer break of two- to three weeks.

For the time of the LS2 shutdown at CERN, it was recommended that CERN and DESY negotiate the possibility to ship one of the EUDET-type pixel beam telescopes currently located at CERN to DESY, so that all three beam lines are equipped with telescopes.

Even without the expected increase of users in 2019/2020, three out of four groups already request a telescope. Supporting the telescopes is of key importance for both the telescope hardware itself and the accompanying software packages. The continued support by DESY would be highly appreciated. The community strongly endorses the continuous development of the telescope hardware as well as the DAQ and reconstruction software. One recent development is the possibility to use the electron beam and the telescope to measure the material budget and composition of various objects with high resolution, which is extremely interesting for many applications. The users suggested to offer this capability as a facility for any user group at the test beam.

The user community also pointed out that it would be highly desirable to have a certain level of coordination between the European labs providing test beams. This is especially true for extended shutdowns like the upcoming one at CERN as the availability of a test beam is vital for detector development in general.

## R-Weg Upgrade

A possibility to meet the requirements expressed by the user community in terms of high-rate and high intensity beams and the production of secondary particles would be to use the current R-Weg as a fourth beam line. The R-Weg was the former transfer line from DESY II to DORIS and the very first part is currently used as a dump for the DESY II beam. It could be relatively easily revived as a fourth beam line by extracting the DESY II beam into this beam line. The primary DESY II bunches could be used, providing 6.3 GeV electrons with intensities of up to $10^{10}$ electrons. In practice, the primary beam would have to be reduced in intensity and broadened in order to offer rates which are useful for detector tests. The extracted beam could be sent to a secondary target to produce pions and muons with a few GeV.

Given the timescales also for the HL-LHC detector upgrades, realizing this upgrade in the next few years is highly desirable.

## Test Beams at DESY 2025+

Given the timescales involved and the unknown scope of future projects, it is difficult to estimate the precise needs for test beams a decade into the future. There is a clear need for test beams due to their central role for detector development in particle physics, nuclear physics, photon science and beyond. Under the assumption that PETRA IV is being realized and that a new injector (currently named DESY IV) is required, the future DESY IV synchrotron should be able to provide test beams at least in a similar fashion as DESY II does today. An important extension highly favored by the user community is the possibility to have resonant extraction in DESY IV. This would allow to provide test beams at the highest energies possible with DESY IV.

## A resource for outreach

Summer student projects have been taking place at the DESY II Test Beam Facility since several years and have been very popular with the students. Since two years teachers from schools from all over Germany have had the opportunity to do experiments at the Test Beam as part of a "Particle Physics for Teachers" week-long seminar at DESY. Both programs will be continued in the future. It was proposed to offer high school students aged 16 or older the possibility to propose and conduct experiments at the DESY II Test Beam Facility similar to what is already done at the Beamlines4Schools programme [5] at CERN. This will be further pursued in the upcoming years.

## Conclusion

The test beam at DESY is of great importance to a broad community and essential for detector development worldwide. The international community, mostly from particle and nuclear physics, has expressed a strong need for this facility. The overall assessment of the community is that the facility is running very well, the availability of beams are excellent and the instrumentation available to the users is key ingredient for its success. The test beam is also a great resource for training the next generation of students. It is equally possible to engage with teachers and potentially high-school students and promote interest in physics in general and with particle physics and detectors in particular.

Having higher particle rates available is the most desirable improvement, both in terms of overall rate as well as in terms of particles per bunch. This is closely linked with having a bunch train mode to more closely emulate conditions at colliders like the LHC or the ILC.

The possibility of having secondary particles like pions and muons is very attractive as well. The R-Weg upgrade is one way to meet the needs of the user community.

The instrumentation provided should develop in line with the future performance of the beam lines. A major upgrade of the telescopes in the time frame of 2020+ to keep up with the improved performance of the beam lines and the increasing user requirements should be considered.

The need for test beams - while difficult to exactly specify for a decade in advance - will not diminish after the completion of the HL-LHC upgrades. First of all, the use of test beams will continue after installation to improve understanding the upgraded detector systems and detector development for future facilities will continue to require high-quality test beams at DESY for many years to come.